\documentclass[authoryear,final,5p,times,twocolumn]{elsarticle}

\usepackage{lineno}
\usepackage{graphicx,amssymb,wasysym,natbib}
\usepackage[ansinew]{inputenc}
\usepackage[percent]{overpic}
\usepackage{color}

\journal{Icarus}

\begin{document}
\newcommand{\actaa}{Acta Astronomica}
\newcommand{\ao}{Applied Optics}
\newcommand{\aap}{Astronomy and Astrophysics}
\newcommand{\aapr}{Astronomy and Astrophysics Reviews}
\newcommand{\aj}{The Astronomical Journal}
\newcommand{\apj}{The Astrophysical Journal}
\newcommand{\apjl}{The Astrophysical Journal, Letters to the Editor}
\newcommand{\apjs}{The Astrophysical Journal, Supplement Series}
\newcommand{\aplett}{Astrophysics Letters and Communications}
\newcommand{\apspr}{Astrophysics Space Physics Research}
\newcommand{\apss}{Astrophysics and Space Science}
\newcommand{\aaps}{Astrophysics and Space Science, Supplement Series}
\newcommand{\araa}{Annual Review of Astronomy and Astrophysics}
\newcommand{\gca}{Geochimica Cosmochimica Acta}
\newcommand{\grl}{Geophysical Research Letters}
\newcommand{\icarus}{Icarus}
\newcommand{\mnras}{The Monthly Notices of the Royal Astronomical Society}
\newcommand{\jcp}{Journal of Chemical Physics}
\newcommand{\jcis}{Journal of Colloid and Interface Science}
\newcommand{\jfm}{Journal of Fluid Mechanics}
\newcommand{\jgr}{Journal of Geophysical Research}
\newcommand{\planss}{Planetary and Space Science}
\newcommand{\nat}{Nature}
\newcommand{\prl}{Physical Review Letters}
\newcommand{\pra}{Physical Review A}
\newcommand{\prb}{Physical Review B}
\newcommand{\prc}{Physical Review C}
\newcommand{\prd}{Physical Review D}
\newcommand{\pre}{Physical Review E}
\newcommand{\solphys}{Solar Physics}
\newcommand{\ssr}{Space Science Reviews}
\newcommand{\zap}{Zeitschrigt f{\"u}r Astronomie} 

\begin{frontmatter}

\title{Microgravity experiments on the collisional behavior of Saturnian ring particles}

\author[tubs]{Daniel Hei\ss elmann\corref{cor1}}
\ead{d.heisselmann@tu-bs.de}
\address[tubs]{Institut f\"ur Geophysik und extraterrestrische Physik,
Technische Universit\"ut Braunschweig, Mendelssohnstra\ss e 3, 38106 Braunschweig, Germany}
\author[tubs]{J\"urgen Blum}
\author[hjf]{Helen J.~Fraser}
\address[hjf]{Department of Physics, University of Strathclyde, SUPA (Scottish Universities Physics Alliance), John Anderson Building, 107 Rottenrow East, Glasgow G4 0NG, United Kingdom}
\author[tubs]{Kristin Wolling}
\cortext[cor1]{Corresponding author}

\begin{abstract}
In this paper we present results of two novel experimental methods to investigate the collisional behavior of individual macroscopic icy bodies. The experiments reported here were conducted in the microgravity environments of parabolic flights and the Bremen drop tower facility. Using a cryogenic parabolic-flight setup, we were able to capture 41 near-central collisions of 1.5-cm-sized ice spheres at relative velocities between $6$ and $22\,\mathrm{cm\,s^{-1}}$. The analysis of the image sequences provides a uniform distribution of coefficients of restitution with a mean value of $\overline{\varepsilon} = 0.45$ and values ranging from $\varepsilon = 0.06$ to $0.84$. Additionally, we designed a prototype drop tower experiment for collisions within an ensemble of up to one hundred cm-sized projectiles and performed the first experiments with solid glass beads. We were able to statistically analyze the development of the kinetic energy of the entire system, which can be well explained by assuming a granular `fluid' following Haff's law with a constant coefficient of restitution of $\varepsilon = 0.64$. We could also show that the setup is suitable for studying collisions at velocities of $< 5\,\mathrm{mm\,s^{-1}}$ appropriate for collisions between particles in Saturn's dense main rings.
\end{abstract}


\end{frontmatter}


\section{Introduction}\label{s_intro}

Saturn's rings are one of the most tremendous astronomical objects in our solar system, and have been studied for centuries by ground-based astronomy (G. Galilei, J.~D. Cassini). More recently, deep-space missions (e.g. \emph{Voyager} and \emph{Cassini}) have allowed for in-depth investigations of these structures. Today it is
clear that the dense main rings (A, B and C ring and Cassini
Division) consist of myriads of particles orbiting Saturn in Keplerian orbits creating a disk only a few meters thick
\citep{tiscareno_et_al2007Icarus, hedman_et_al2007AJ, porco_et_al2008AJ}. Spectroscopic
observations \citep{poulet_et_al2003A&A, nicholson_et_al2008Icarus} and radio occultation data \citep{tyler_et_al1981Science, marouf_et_al1983Icarus, zebker_et_al1985Icarus}
have confirmed that the rings consist predominantly of bodies of almost pure water ice, with sizes between $1\,\mathrm{cm}$ and
$10\,\mathrm{m}$. Many of the observed ring features are caused by
interactions with nearby moons and moonlets which
gravitationally perturb their trajectories and thus increase their
orbital eccentricity and the ring thickness. Counteracting this, frequent inelastic collisions at very low relative velocities \citep[$v\lesssim 0.5\,\mathrm{cm\,s^{-1}}$, see review by][]{esposito2002RPPh} are means to dissipate kinetic
energy and thereby re-circularize the orbits and confine the
height of the ring plane.\par

Several theoretical and
numerical studies have been carried out to investigate how structures inside the dense rings are created,
how ring-particle orbits evolve over time and the expected outcome of typical collisions between ring particles. Kinetic theories were first developed by \citet{goldreich_tremaine1978Icarus} and
\citet{hameen-anttila1978Ap&SS}, based on earlier work by
\citet{trulsen1971Ap&SS, trulsen1972Ap&SS}. These theories, which
were applicable to thin granular gases of a single species (i.e. one
particle size), were extended to dense granular disks
\citep{borderies_et_al1985Icarus, araki_tremaine1986Icarus,
araki1991Icarus}, and subsequently explored with N-body simulations
\citep[e.g.][]{wisdom_tremaine1988AJ, salo1991aIcarus, salo1991bIcarus} that also included the particle size distribution described by
\citet{zebker_et_al1985Icarus}. In addition, visco-elastic
collision models, developed by \citet{spahn_et_al1995CSF}
and \citet{brilliantov_et_al1996PRE}, have been extended to also treat fragmentation and coagulation
\citep{spahn_et_al2004EL, albers_spahn2006Icarus}. The models of
\citet{schmidt_et_al2001Icarus} and \citet{salo_et_al2001Icarus}
show that the amount of energy dissipated through collisions
determines the stability of homogeneous dense rings, and thus is a
crucial parameter in understanding the formation of structures, instabilities
and wakes as observed in Saturn's rings \citep{colwell_et_al2006GRL, colwell_et_al2007Icarus, hedman_et_al2007AJ, thomson_et_al2007GRL}.\par

In most N-body simulations and kinetic theories, the
inelastic collisional behavior is described by a single parameter, the coefficient of restitution $\varepsilon$, which is the ratio of the relative
velocities $v'$ after the encounter to the relative velocities $v$ before the encounter, and is a
measure of the dissipation of the kinetic energy. Although ice-paricle collisions play a vital role in determining the stability and evolution of Saturn's rings, as well as in the formation of icy planetesimals in protoplanetary disks, their collision properties have hardly been investigated in the low velocity regime. Previous experiments carried out by \citet{bridges_et_al1984Nature} and
\citet{hatzes_et_al1988MNRAS}, used a disk pendulum to study
central impacts of large ice bodies onto a flat solid ice target, and
showed a decrease in the coefficient of restitution with
decreasing impact velocity and increasing amount of surface frost.
The experiments were extended to include grazing collisions of polished
ice spheres \citep{supulver_et_al1995Icarus}. \citet{higa_et_al1996P&SS, higa_et_al1998Icarus} conducted impact
experiments of different-sized ice spheres ($r=0.14 - 3.6\,\mathrm{cm}$) into a larger ice target. They found that the
coefficient of restitution decreases with increasing particle
size and is independent of temperature.\par

A major limitation of all these experimental studies
\citep{bridges_et_al1984Nature, hatzes_et_al1988MNRAS,
higa_et_al1996P&SS, higa_et_al1998Icarus} is that they only treat
impacts into a large body of infinite mass and infinitesimal
surface curvature. In the case of the pendulum
experiments an additional problem is that the mass and moment of inertia of the ice sphere
are not realistic as they reflect those of the
sample-pendulum-system instead of that of the individual
projectile. However, the results obtained by \citet{bridges_et_al1984Nature} are commonly used in numerical simulations of Saturn's rings.\par

In this work we present two novel experimental methods for
investigating collisions of Saturnian ring particles. In Sect.
\ref{s_setup} the two experimental setups for collisions of
individual ice particles and ensembles of particles are described.
The results of our microgravity experiment campaigns are presented in
Sect. \ref{s_results}, and their application to Saturn's dense
rings is given in Sect. \ref{s_conclusion}.

\section{Experimental setup}\label{s_setup}
For our microgravity studies we used two different experimental
approaches. The parabolic flight setup
was designed to study collisions of pairs of individual ice projectiles, whereas the Bremen drop tower facility experiment is a prototype, built to study a large number
of low-velocity collisions within an ensemble of particles.\par

Microgravity is required to be able to investigate low-velocity
pair-collisions of free particles with an arbitrary impact parameter
(i.e. from central to grazing collisions), because it is the only way to access all the translational and rotational degrees of freedom.
Parabolic flights offer typically 22 s of weightlessness per
parabola with a residual acceleration of a few times 0.01 $g_0$,
where $g_0$ is Earth's gravitational acceleration. Drop-tower
experiments achieve residual accelerations of better than
$10^{-5}~ g_0$ for up to 9~s.

\subsection{Parabolic flight experiment}\label{ss_pf_exp}
Measurements were conducted on the German Space Agency's (DLR)
$12^{\mathrm{th}}$ Parabolic Flight Campaign, using a setup
consisting of a cryogenic cooled experiment body, placed inside a high-vacuum
chamber. The experimental apparatus is
composed of a massive copper block which acts as a thermal reservoir, with a
rotating sample repository covered by a copper shield, which plays both a protective and cryo-cooling role.
The entire system is initially cooled to roughly 77~K with liquid nitrogen ($\mathrm{N_2}$)
spiraling through an attached pipe system. Up to 34 pairs of cm-sized ice particles can be stored in the sample repository, which allows us to collide approximately 1 pair per parabola. To accelerate the ice particles into the collision volume, two diametrically opposed pistons are used, driven by two synchronized DC motors, which accelerate the particles to maximum velocities between $3-17\,\mathrm{cm\,s^{-1}}$. These are positioned outside the
cryo-vacuum chamber, and enter the vacuum by differentially pumped feedthroughs. Since the projectiles are driven towards each
other from exactly opposite directions, the resulting binary
collisions are predominantly centralized, with normalized impact
parameters, $b/R$, close to zero, where $b$ is the center-of-mass
distance of the two projectiles perpendicular to the velocity
vector, and $R$ is sum of the particles' radii. The collisions are
recorded with a high-speed, high-resolution digital recording system,
operating at 107 frames per second (fps) and equipped with
beam-splitter optics to provide full three-dimensional collision
information (i.e. impact parameter and relative velocity). For a
more detailed technical description please refer to \citet{salter_et_al2009RSI}.\par

41 image sequences of colliding ice spheres ($15\,\mathrm{mm}$ diameter) were recorded during  DLR's $12^{\mathrm{th}}$ Parabolic Flight Campaign in April 2008. The samples were produced by freezing purified water inside silicone molds in a regular kitchen freezer, to produce $15\,\mathrm{mm}$-sized, hexagonal-ice spheres. After removing the molds from the freezer, the spheres were extracted and stored inside a bath of liquid $\mathrm{N_2}$ from where they were individually loaded into the sample
repository, which had been pre-cooled, and cooled further to temperatures of $\sim
80\,\mathrm{K}$. This procedure is required to minimize the amount of frost on the samples' surfaces. Although the maximum exposure time of the samples to the humid air was less than one minute, some experiments show tiny amounts of surface material (which might be frost) chipped off during the collisions.\par

Since liquid $\mathrm{N_2}$ is prohibited on board the parabolic-flight aircraft, the cooling
procedure had to be stopped before take-off. Consequently, the thermal block warmed slightly prior to take-off, and subsequently during flight, resulting in sample temperatures ranging from $130-180\,\mathrm{K}$.\par

\subsection{Drop tower experiment}\label{ss_dt_exp}
\begin{figure}[!tb]
    \center
    \includegraphics[width=21pc]{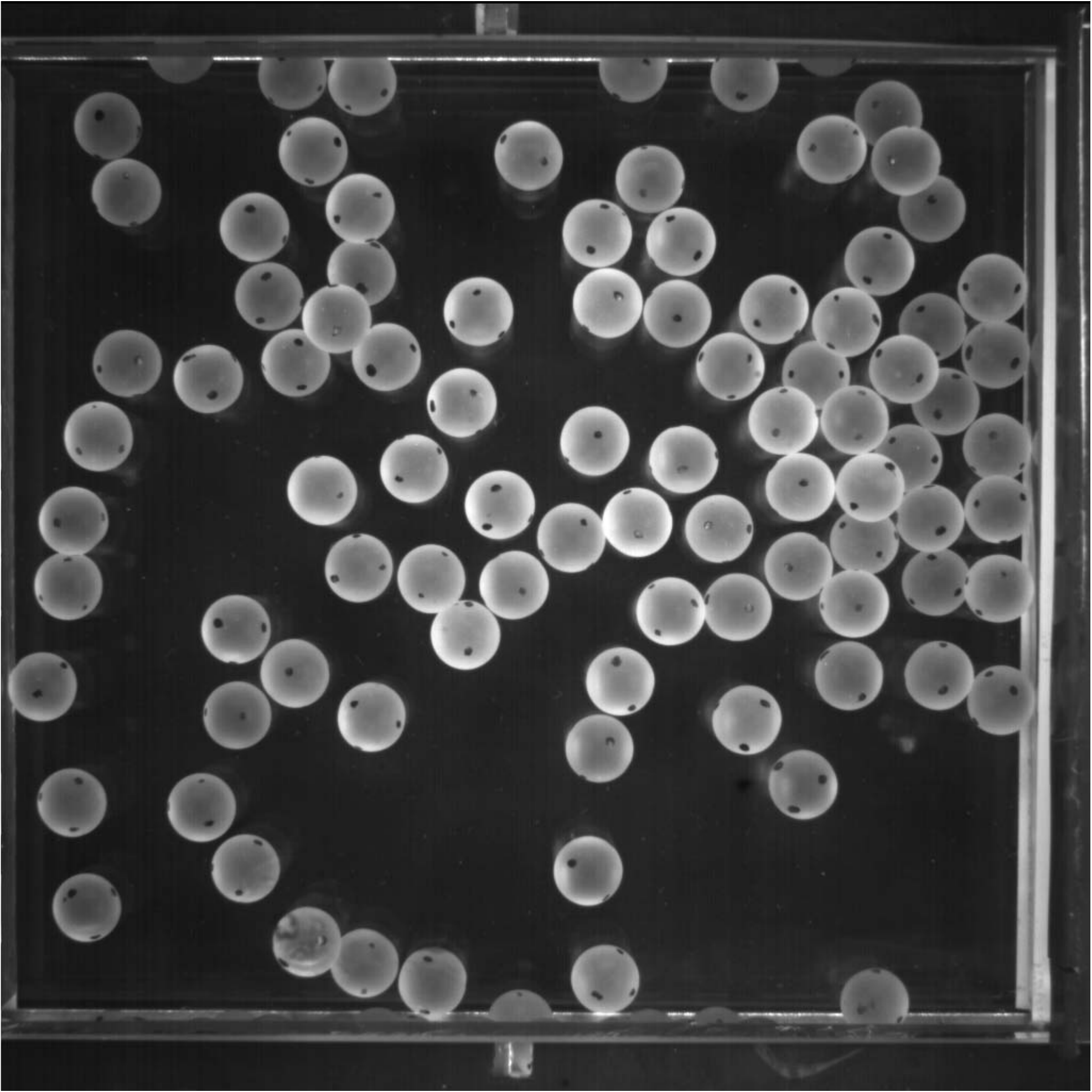}
    \caption{Single image of an ensemble of cm-sized glass spheres colliding in microgravity. The black dots on the particles are marks for later use in the determination of particle rotation (see additional online material).}
    \label{f_dt_exp}
\end{figure}

A quasi-two-dimensional rectangular glass box of $150\times
150\times 15\,\mathrm{mm^{3}}$ volume (see Fig. \ref{f_dt_exp}) was constructed to investigate a large number of collisions within an ensemble of $\sim 100$ (cm-sized) particles.
The test-chamber's walls are made of 2~cm thick glass to ensure
highly elastic collisions between the sample particles and the walls. Two sets of
up to 50 projectiles, stored opposite each other, were accelerated into the test chamber with an initial velocity of
$\sim 10\,\mathrm{cm\,s^{-1}}$ by two motor-driven glass bars
that afterwards seal these entry holes (Fig. \ref{f_dt_exp}). During the 9~s of
microgravity\footnote{The use of the Bremen drop tower's catapult
facility almost doubled the duration of microgravity compared to a regular drop without using the catapult.}, the
particle positions were recorded by a high-speed, high-resolution
camera operating at 115~fps, and two overview cameras (25~fps) positioned at an angle of
$16.25\textdegree$ relative to the chamber, to enable three-dimensional trajectory reconstruction.

Here we focus on the results of one out of seven microgravity
experiments conducted at the Bremen drop tower in April 2008. This was a prototype experiment for future investigations
that will be equipped with cryogenic facilities and therefore use ice particles. However in this case, the experiment was performed with 92
spherical glass samples of $10\,\mathrm{mm}$ diameter, with a
slightly roughened surface. At the beginning of the microgravity
period, two sets of 32 particles were injected into the test
chamber where 28 samples were at rest. The experiment chamber was
not evacuated, but the influence of the air drag acting on a
sample particle does not significantly affect its velocity. A smooth glass sphere (10~mm diameter) has a Reynolds number $Re\approx 1.6$ at a velocity
of $v=4\,\mathrm{mm\,s^{-1}}$, and $Re\approx 62$ at a velocity of
$v=10\,\mathrm{cm\,s^{-1}}$. From this the relative velocity loss $\Delta v/v$ due to air
drag between two consecutive collisions can be estimated to be less
than a few times $10^{-3}$.\par

Standard methods of image processing and particle
tracking were used to obtain the particles' positions from the recorded image sequence. A statistical analysis of all particles' velocities as a function of time was performed, which means that -- although generally possible -- we did not treat individual collisions but concentrated on the velocity development of the entire ensemble.
In future experiments, the individual collisions, their impact
parameters and the excitation of rotational motion will be
investigated. The results of our analyses can be found in Sect.
\ref{ss_dt_res}.

\section{Results}\label{s_results}

\subsection{Parabolic flight experiment}\label{ss_pf_res}
\begin{figure}[!tb]
    \center
    \includegraphics[width=21pc]{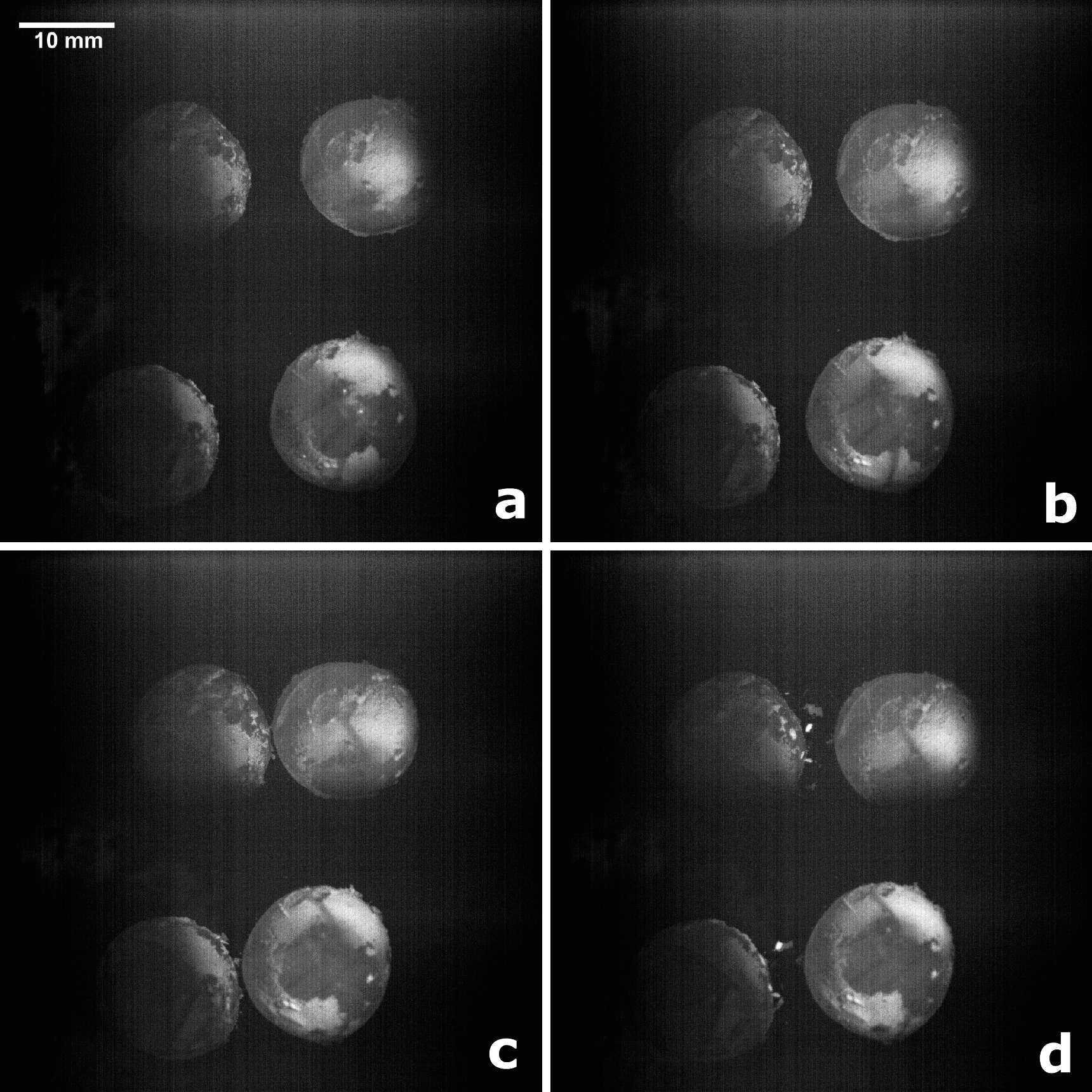}
    \caption{Image sequence of two 15-mm-sized ice spheres colliding at a relative velocity of $14\,\mathrm{cm\,s^{-1}}$. The temporal separation between two consecutive frames is $\frac{4}{107}\,\mathrm{s}$. The images were captured using a beam-splitter optics. Therefore, each image frame contains two views of the set of particles, which are separated by an angle of $48.8\textdegree$.}
    \label{f_pf_collage}
\end{figure}

The binary collision experiments performed during DLR's $12^{\mathrm{th}}$ Parabolic Flight Campaign occurred in free space and so were truly three-dimensional events. Thus, data from two projections of each event captured using beam-splitter optics (at an angular separation of $48.8\textdegree$) were used to determine the particle motion (see Fig. \ref{f_pf_collage}). The two sets of two-dimensional coordinates were combined with a transformation algorithm to produce the sample coordinates in three dimensions, and the trajectories were determined by linear fits to the data. From this, the relative velocities before and after the collision, $v$ and $v'$ respectively, and the normalized impact parameter, $b/R$, were calculated. Additionally, the velocity components normal and tangential to the colliding surfaces ($v_{\bot}$ and $v_{\|}$, respectively) could be extracted.

\begin{figure}[!tb]
    \center
    \includegraphics[width=21pc]{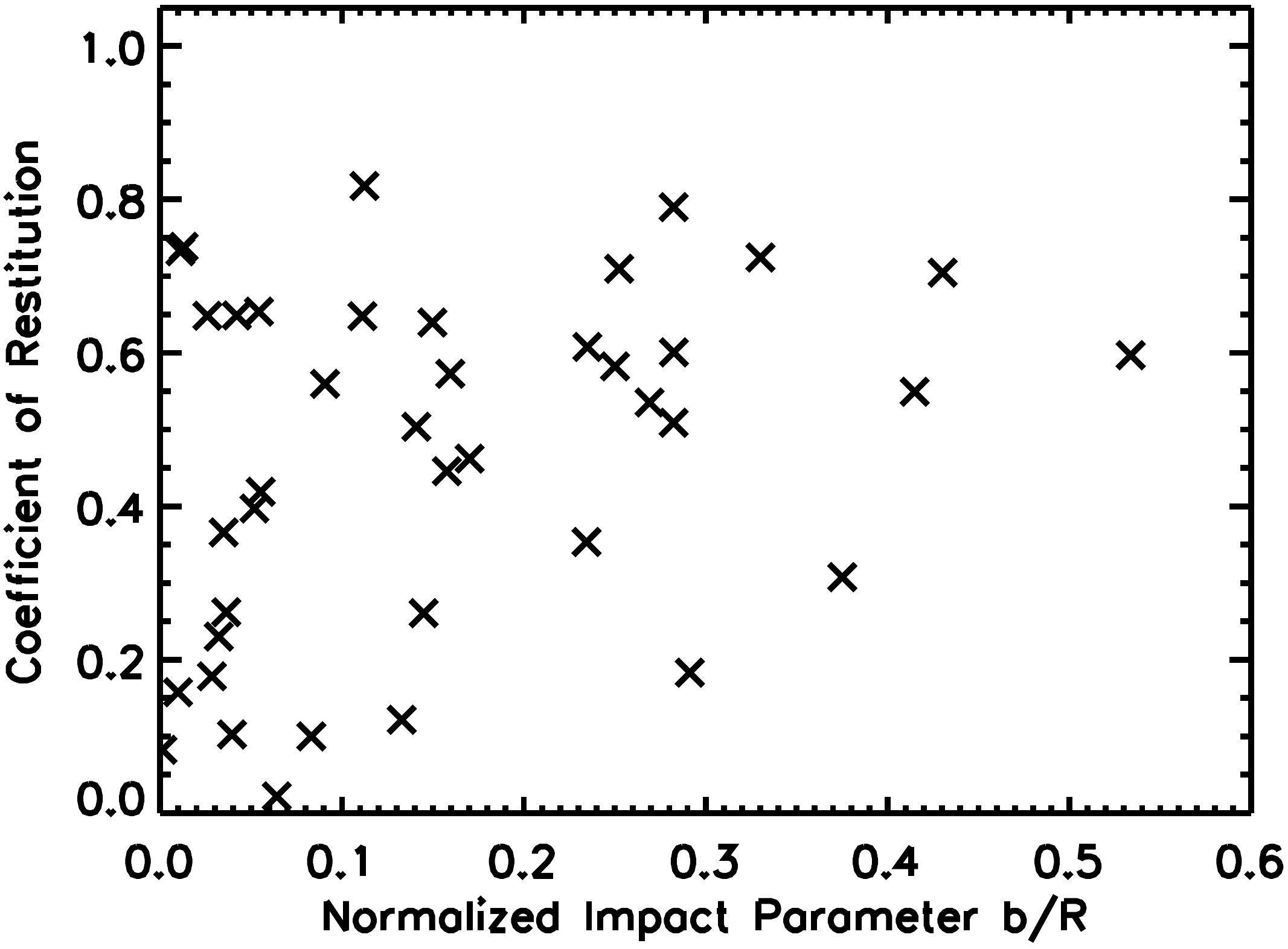}
    \caption{Obtained coefficients of restitution for ice particle collisions at near-central impact parameters in the range $b/R=0 - 0.5$. There is no evidence of a correlation between the two variables.}
    \label{f_pf_res}
\end{figure}

\begin{figure}[!tb]
    \center
    \includegraphics[width=21pc]{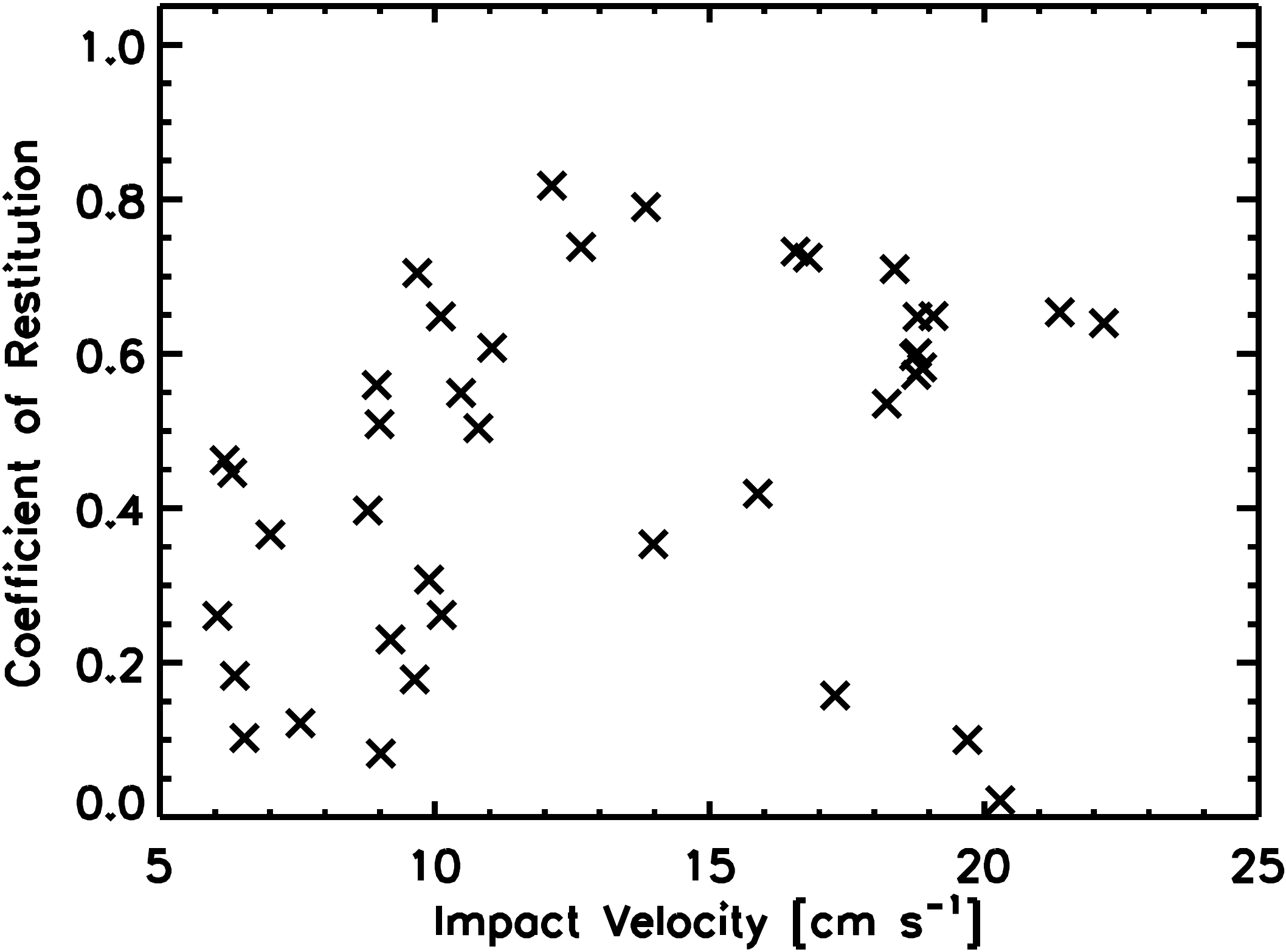}
    \caption{Obtained coefficients of restitution for ice particle collisions in the velocity range $v=6 - 22\,\mathrm{cm\,s^{-1}}$. A linear correlation analysis shows no evidence for a correlation between these variables.}
    \label{f_pf_res_v}
\end{figure}

From Fig. \ref{f_pf_res} it is clear that the impact parameters are restricted to near-central collisions (due to the limitations of our experimental setup), ranging from $b/R=0$ to $0.5$ with a mean value $\overline{b/R}=0.16$.  However, simple geometrical considerations show that in reality glancing collisions are the most frequent ones. Thus the captured collisions with impact parameters smaller than $b/R = 0.5$ cover only a fraction of $P(i.\,p. < 0.5)=(b/R)^2=0.25$ of the statistically occurring impacts in Saturn's rings. From the observed impact parameters and coefficients of restitution no evidence for a correlation between these parameters could be found by linear correlation analysis, in which we computed the linear correlation coefficient:

\begin{eqnarray}
  r_{b/R}=\frac{\sum_{i=0}^n \left(\varepsilon _i-\overline{\varepsilon}\right)\left(\frac{b_i}{R_i}-\overline{\frac{b}{R}}\right)}{\sqrt{\sum_{i=0}^n \left(\varepsilon _i-\overline{\varepsilon}\right)^2} \cdot \sqrt{ \sum_{i=0}^n \left(\frac{b_i}{R_i}-\overline{\frac{b}{R}}\right)^2}}\;,
\end{eqnarray}
\noindent
where $r _{b/R} ^2=0.10$ gives the fraction of the data which can be explained by a linear dependence. Based on a similar analysis of the coefficient of restitution $\varepsilon$ and the relative impact velocity $v$ ($r _{v} ^2=0.11$; see Fig. \ref{f_pf_res_v}) a correlation of those can also be ruled out. In a detailed analysis of the coefficient of restitution in normal ($\varepsilon _{\bot}$) and tangential direction ($\varepsilon _{\|}$) no evidence for a correlation of these values with the impact parameter or the impact velocity could be found. Elaborate investigation showed that also a dependence of the coefficient of restitution on a combination of the collision velocity and the impact parameter could be ruled out.


\begin{figure}[!tb]
    \center
    \includegraphics[width=21pc]{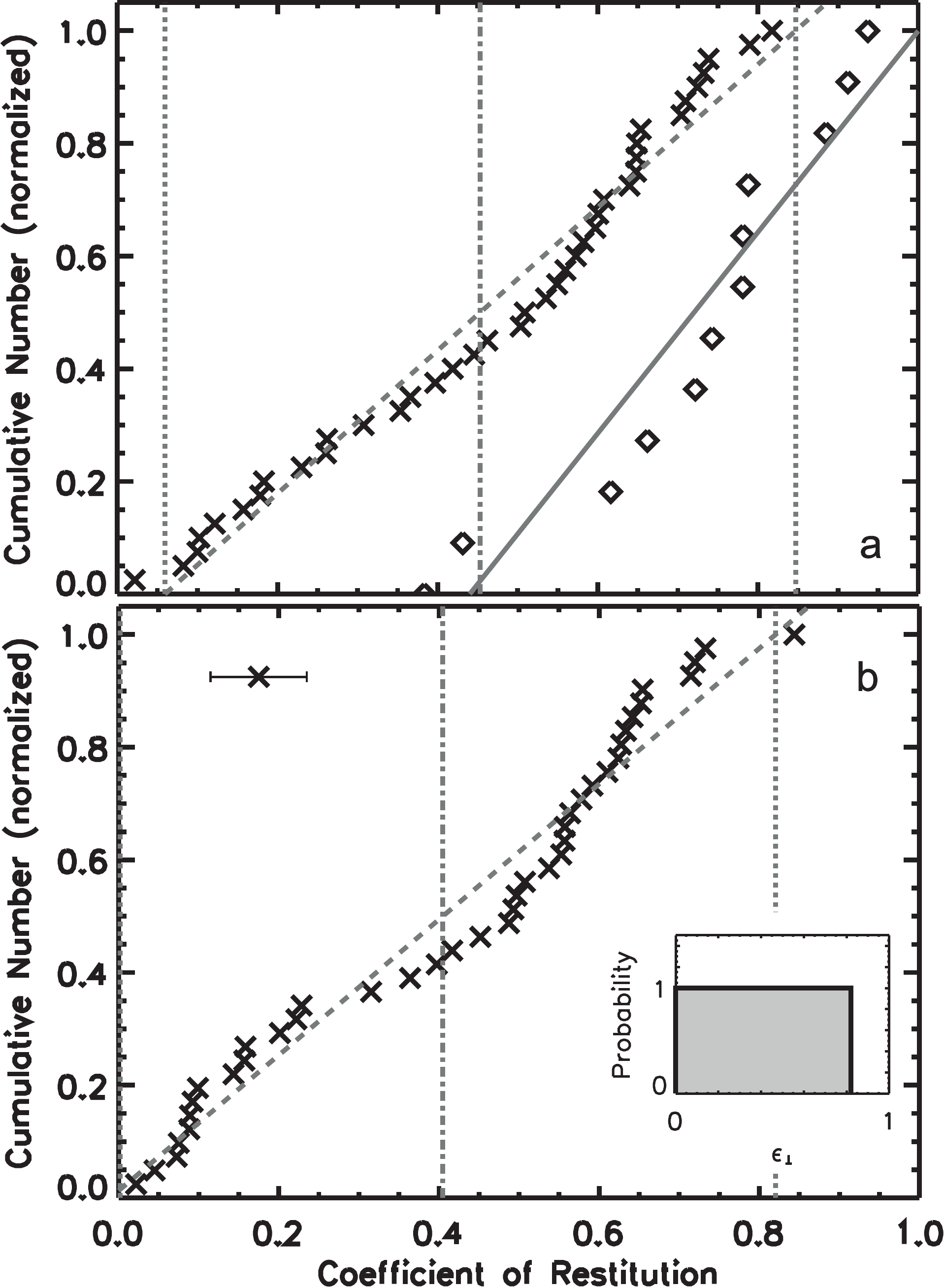}
    \caption{(a) Normalized cumulative number of ice particle collisions with coefficients of restitution $\le \varepsilon$ for all captured collisions in the parabolic-flight experiments (crosses). The distribution of $\varepsilon$ can be fitted by a uniform distribution which is given by the dashed curve. The mean coefficient of restitution is at $\overline{\varepsilon}=0.45$ (denoted by the dash-dotted line) with an overall range of $\varepsilon =0.06$ to $0.84$ (dotted lines). The diamonds represent the data obtained from binary collisions of glass beads in the laboratory mini drop tower. The coefficients of restitution span a range of $\varepsilon \approx 0.35$ to $0.95$ (solid curve). (b) Normalized cumulative number of collisions with normal coefficients of restitution $\le \varepsilon_{\bot}$ for all 41 captured experiments. The normal component has a mean value of $\overline{\varepsilon_{\bot}} = 0.41 \pm 0.04$ with a standard deviation of individual measurements of 0.24. The measurement error of an individual measurement is 0.06 (indicated by the error bars). The insert shows a schematic distribution of coefficients of restitution that could be included into numerical simulations.}
    \label{f_pf_cumulative}
\end{figure}

Hence, we can treat the coefficient of restitution data in a statistical way. One way to do this, is to plot the cumulative number of collisions with a coefficient of restitution $\le \varepsilon$ (Fig. \ref{f_pf_cumulative}a). It turns out that this can be fitted by a uniform distribution (see dashed curve in Fig. \ref{f_pf_cumulative}a). The fit provides a mean coefficient of restitution of $\overline{\varepsilon} = 0.45$ and an overall range of $\varepsilon =0.06 - 0.84$ (dotted lines in Fig. \ref{f_pf_cumulative}a).\par

The obtained normal coefficient of restitution $\varepsilon _{\bot}$ was treated in a similar way. Fig. \ref{f_pf_cumulative}b shows that values of $\varepsilon _{\bot}$ fall in the interval $0\ldots0.82$ (dashed lines). The analysis of our dataset yields a mean value $\overline{\varepsilon _{\bot}}=0.41$ with  an error of the mean $\sigma_{\overline{\varepsilon_{\bot}}}=0.04$ and individual measurement errors of $\Delta \varepsilon_{\bot}=0.06$. The standard deviation of the individual measurements from the mean value was calculated to $\sigma_{\varepsilon_{\bot}}=0.24$ indicating a significant scatter of the values for $\varepsilon _{\bot}$ around $\overline{\varepsilon _{\bot}}$ which is not caused by the measurement uncertainties. We therefore recommend to include a flat statistical distribution of coefficients of restitution (see insert in Fig. \ref{f_pf_cumulative}b) rather than a single value to the models investigating Saturn's rings. However, it is likely that the wide range of coefficients of restitution observed in our parabolic flight experiments reflects the anisotropic nature of the ice samples' surface properties, especially the surface roughness or previously slightly melted areas altering the collisional properties.\par

For the analysis of the tangential component $\varepsilon _{\|}$ only experiments with tangential velocities $v_{\|}\geq 5\,\mathrm{mm\,s^{-1}}$ (36 out of 41 experiments) were taken into account to avoid $\varepsilon _{\|}=v^\prime_{\|}/v_{\|}$ approaching unrealistically large values when the denominator $v_{\|}$ approaches zero. The mean value of the tangential coefficient of restitution was measured to be $\overline{\varepsilon _{\|}}=1.08$ with an error of $\sigma_{\overline{\varepsilon_{\|}}}=0.19$ and individual standard deviation of $\sigma_{\varepsilon_{\|}}=1.11$, which is in fair agreement with the value found by \citet{supulver_et_al1995Icarus}. The large individual measurement errors of $\Delta\varepsilon _{\|}=0.38$ (due to the small absolute values of $v_{\bot}$) do not allow any secure interpretation, however. Table \ref{t_epsilon} summarizes our experimental results for the normal and tangential coefficient of restitution of binary collisions between cm-sized ice spheres.\par

\begin{table}
    \scriptsize
    \caption{Summary of the experimental results of the coefficient of restitution of low-velocity ice-particle collisions. The velocity range was $v_{\bot}=4\ldots22\,\mathrm{cm\,s^{-1}}$ and $v_{\|}=0.5\ldots8\,\mathrm{cm\,s^{-1}}$, respectively.}\vspace{1mm}
    \label{t_epsilon}
    \begin{tabular*}{18.5pc}{cccc|cccc}
        \hline
        \multicolumn{4}{c|}{$\varepsilon_{\bot}$}&\multicolumn{4}{c}{$\varepsilon_{\|}$}\\
        \hline
        $\overline{\varepsilon_{\bot}}$ & $\sigma_{\overline{\varepsilon_{\bot}}}$ & $\sigma_{\varepsilon_{\bot}}$ & $\Delta \varepsilon_{\bot}$&$\overline{\varepsilon_{\|}}$ & $\sigma_{\overline{\varepsilon_{\|}}}$ & $\sigma_{\varepsilon_{\|}}$ & $\Delta \varepsilon_{\|}$\\
        \hline
        0.41 & 0.04 &  0.24 & 0.06&1.08 & 0.19 &  1.11 & 0.38\\
        \hline
    \end{tabular*}
    \vspace{1mm}

    $\overline{\varepsilon_{\bot}},\,\overline{\varepsilon_{\|}}$: mean normal / tangential coefficient of restitution\\
    $\sigma_{\overline{\varepsilon_{\bot}}},\,\sigma_{\overline{\varepsilon_{\|}}}$: error of mean normal / tangential coefficient of restitution\\
    $\sigma_{\varepsilon_{\bot}},\,\sigma_{\varepsilon_{\|}}$: standard deviation of individual measurements of normal / tangential coefficient of restitution from mean value\\
    $\Delta\varepsilon_{\bot},\,\Delta\varepsilon_{\|}$: measurement uncertainty of individual measurement of normal / tangential coefficient of restitution\\
\end{table}

\subsection{Drop tower experiment}\label{ss_dt_res}

To yield the positions of all particles in all image frames, the images were convolved with an image of a single sphere as kernel. The probability maxima, which determine the particle positions, could then be detected automatically. The tracking of all particles over time allowed the calculation of the individual particle's mean velocity\footnote{The reader should note that although the term `velocity' generally describes a vectorial quantity we use it for the absolute value of the velocity. In all cases treating the velocity as a vector this is clearly stated.} between consecutive images. The resulting particle velocities as a function of time show that the system equilibrates very quickly within the first 1.5~s. After this injection phase, we observed dissipation of kinetic energy through inelastic collisions and transformation to rotational motion which was visualized by black marks on the surface of the glass beads (see Fig. \ref{f_dt_exp}).\par

\begin{figure}[!tb]
    \center
    \includegraphics[width=21pc]{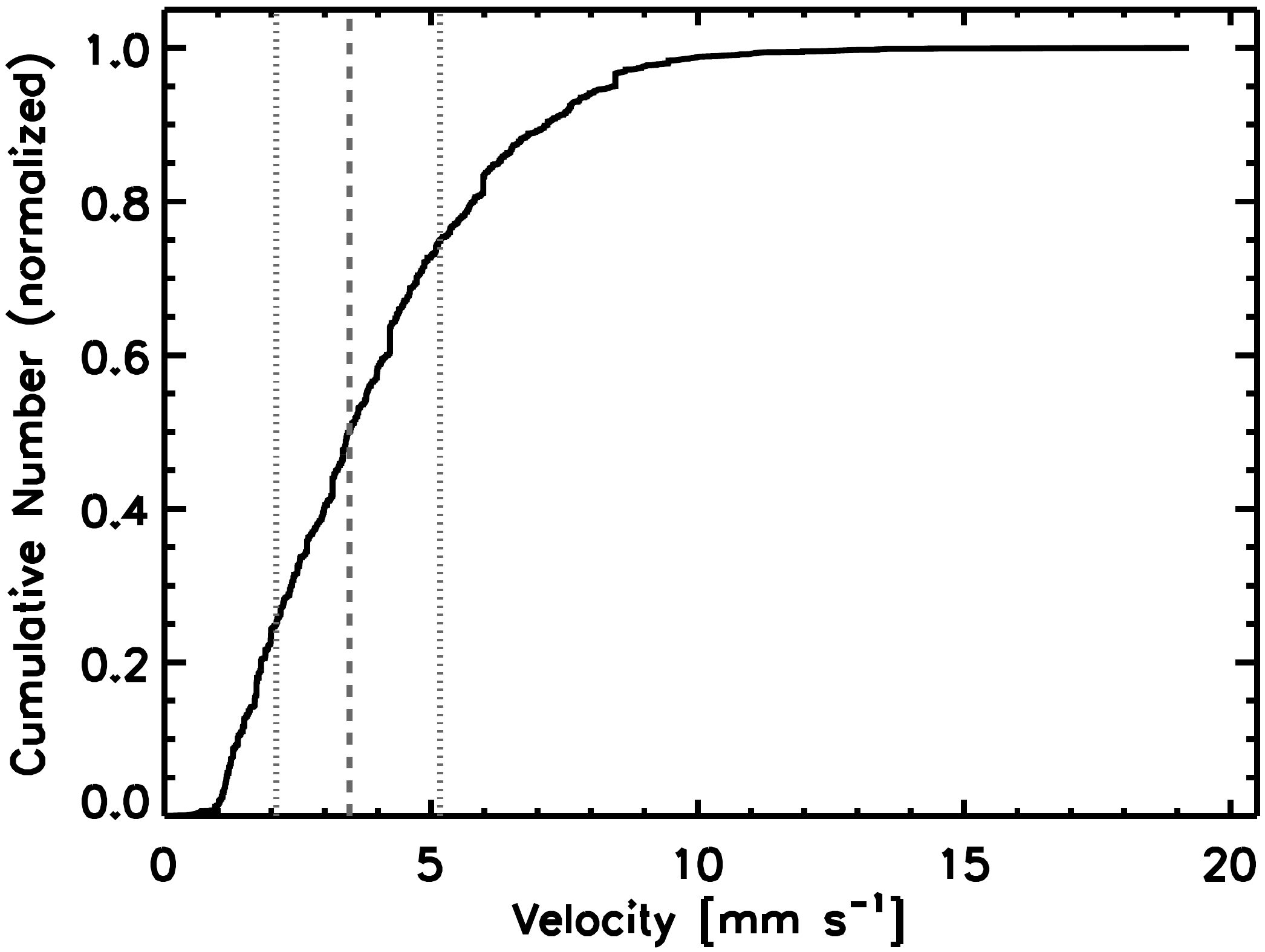}
    \caption{Normalized cumulative number of glass particles with velocity $\le v$ in the time interval of 7 to 8.5~s of microgravity duration for the drop-tower experiments. The dashed line denotes the median particle velocity and the dotted lines indicate the particle velocities covering 50\% of all particles.}
    \label{f_dt_cumulative}
\end{figure}

From the analysis of the particle velocities within the last 1.5~s of experiment time, we obtained a median particle velocity of $3.5\,\mathrm{mm\,s^{-1}}$ and determined that 50 \% of the particles are moving with velocities between $2$ and $5\,\mathrm{mm\,s^{-1}}$ (see dashed and dotted lines in Fig. \ref{f_dt_cumulative}), which corresponds fairly well to the desired velocities for studying collisions in Saturn's dense rings \citep{esposito2002RPPh}.

\begin{figure}[!tb]
    \center
    \includegraphics[width=21pc]{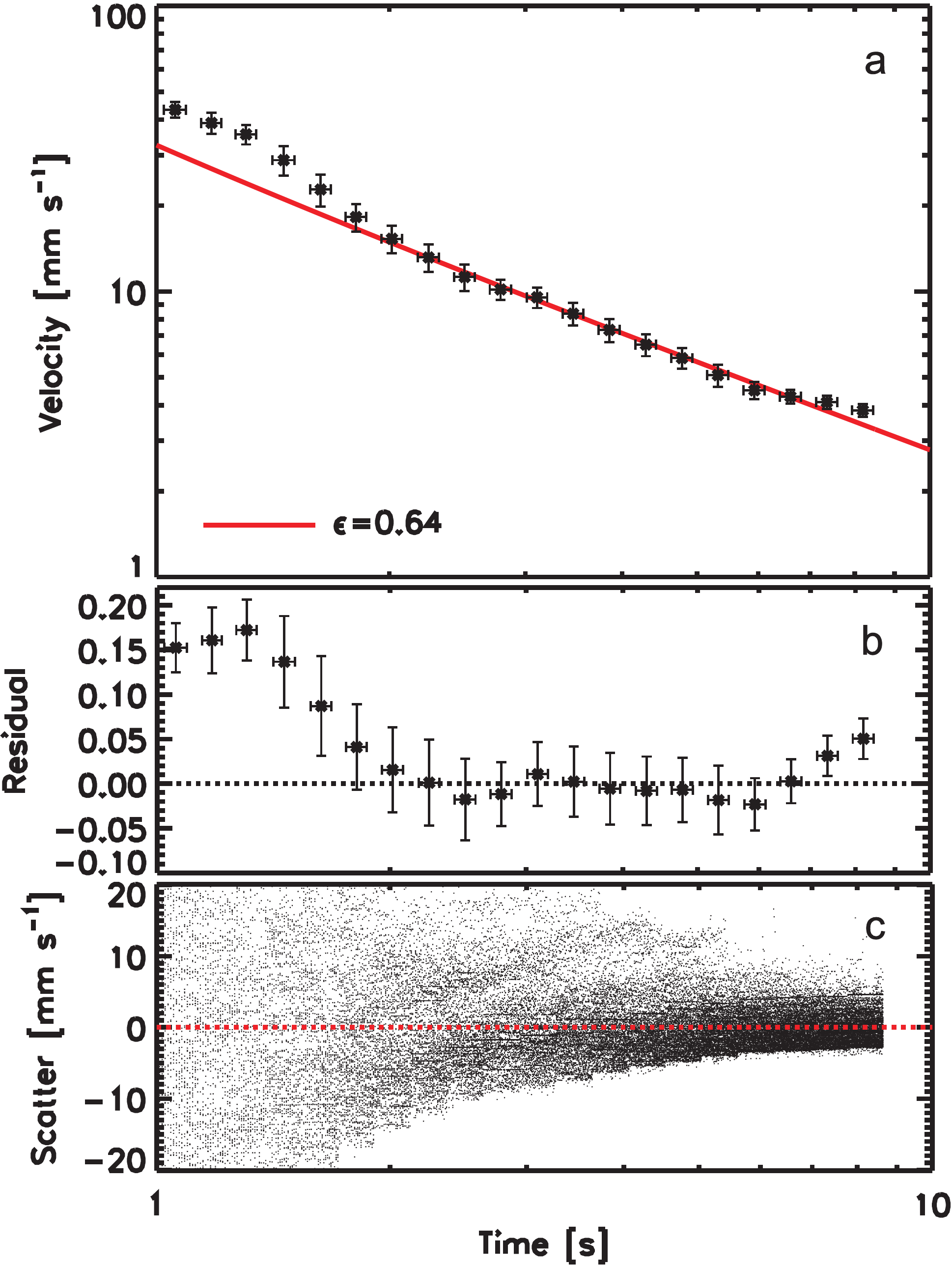}
    \caption{(a) Velocity decrease of the glass particles in the drop-tower experiment due to the dissipation of kinetic energy as a function of experiment duration. The data was binned in logarithmic intervals. It is clearly visible that after a short period of equilibration the dissipation is in agreement with Haff's cooling law for granular gases \citep{haff1983JFM} between $2$ and $\sim 7\,\mathrm{s}$ experiment duration. The solid line denotes the relation given in Eq. \ref{eq_haff} with a coefficient of restitution of $\varepsilon = 0.64$. (b)~The residual between Haff's law and the data shows a good agreement from 2 to 7~s experiment time. The strong deviation within the first 2~s is due to the equilibration phase after the particle injection, whereas the divergence after 7~s can only be explained by a change of the coefficient of restitution towards low relative velocities. (c) The scatter of the individual data points around the fitted curve shows a broad velocity distribution.}
    \label{f_haff}
\end{figure}

The decay of the mean particle velocity of the entire ensemble during the experiment duration is displayed in Fig. \ref{f_haff}a. The individual measurements were logarithmically binned and the mean velocity was computed. The error bars denote the standard deviation of the velocity and experiment time measurements, respectively. Assuming a constant coefficient of restitution $\varepsilon \left(v\right)=const.$, the simple kinetic theory for granular fluids presented by \citet{haff1983JFM} indicates that the temporal rms-velocity evolution can be described by the following function:
\begin{eqnarray}
 v\left(t\right)&=&\frac{1}{\frac{1}{v_0}+(1-\varepsilon )\cdot n\cdot  \sigma \cdot t}\;,\label{eq_haff}
\end{eqnarray}
where $v_0$ is the initial injection velocity of the ensemble, $n$ is the number density and $\sigma = 4\pi r^2$ is the collisional cross-section. Eq. \ref{eq_haff} can be fitted to the binned velocity data, resulting in a coefficient of restitution of $\varepsilon = 0.64$ (see solid line in Fig. \ref{f_haff}a). Thereby, the data represent an average over all impact parameters (i.e. central to grazing impacts) and, thus, collisions are only analyzed statistically, but not as individual binary encounters. The residual of the fit to the data (Fig. \ref{f_haff}b) shows a good agreement of both in the interval from 2 to 7 seconds experiment duration, meaning that the collisional behavior follows Haff's law with $\varepsilon = 0.64$. The strong deviation within the first 2 seconds is due to the equilibration phase of the ensemble after the sample injection, whereas the clear change in slope after $\sim 7$ seconds is a real effect that can either be explained by an increased coefficient of restitution towards low relative velocities or the onset of clustering \citep{goldhirsch_zanetti1993PRL, brito_ernst1998EL, miller_luding2004PRE}. However, the investigation of the latter exceeds the scope of this work.\par

We performed a detailed analysis of individual binary collisions among the constituents of the ensemble in the drop tower experiment as well as of binary collisions of identical glass spheres conducted in a recently built laboratory mini drop tower of 1.5~m height.
In the latter experiments we observed 12 central collisions in the velocity range $1 - 4\,\mathrm{cm\,s^{-1}}$. The measured coefficients of restitution span a range from $\varepsilon \approx 0.35$ to $\varepsilon \approx 0.95$, are randomly distributed (see solid curve in Fig. \ref{f_pf_cumulative}), and show no correlation with impact velocity.\par
\begin{figure}[!tb]
    \center
    \includegraphics[width=21pc]{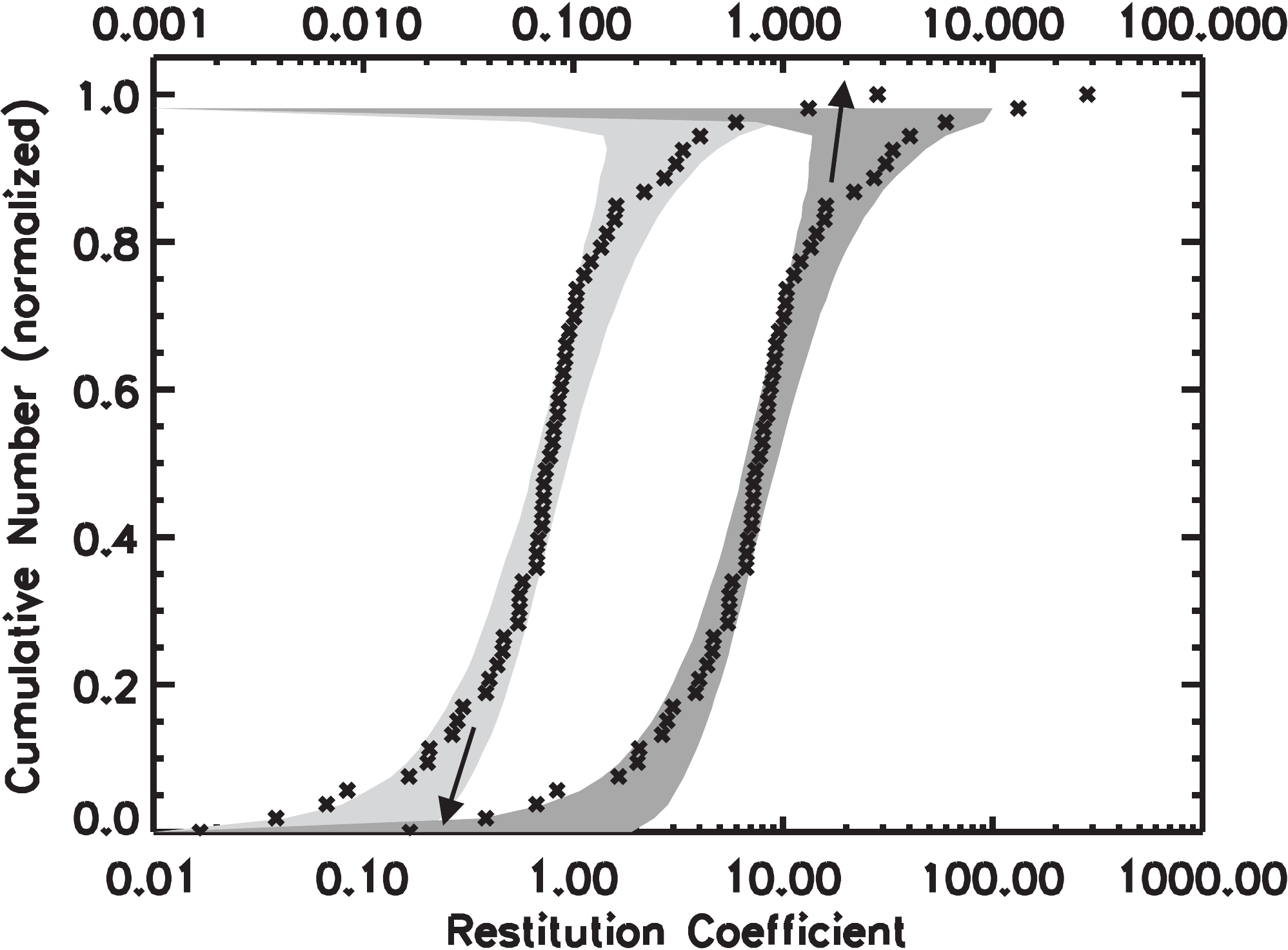}
    \caption{The coefficients of restitution obtained by analysis of the binary collisions of cm-sized glass spheres (crosses) can be fitted by a Maxwell-Boltzmann distribution with a constant coefficient of restitution of $\varepsilon=0.77$ (right shaded area). Using a range of coefficients of restitution ($\varepsilon=0.2\ldots1.0$) results in a very similar velocity distribution function (left shaded area). The shaded areas denote the 2$\sigma$-standard-deviation derived by a Monte Carlo simulation.}
    \label{f_maxwell}
\end{figure}
\begin{figure}[!tb]
    \center
    \includegraphics[width=21pc]{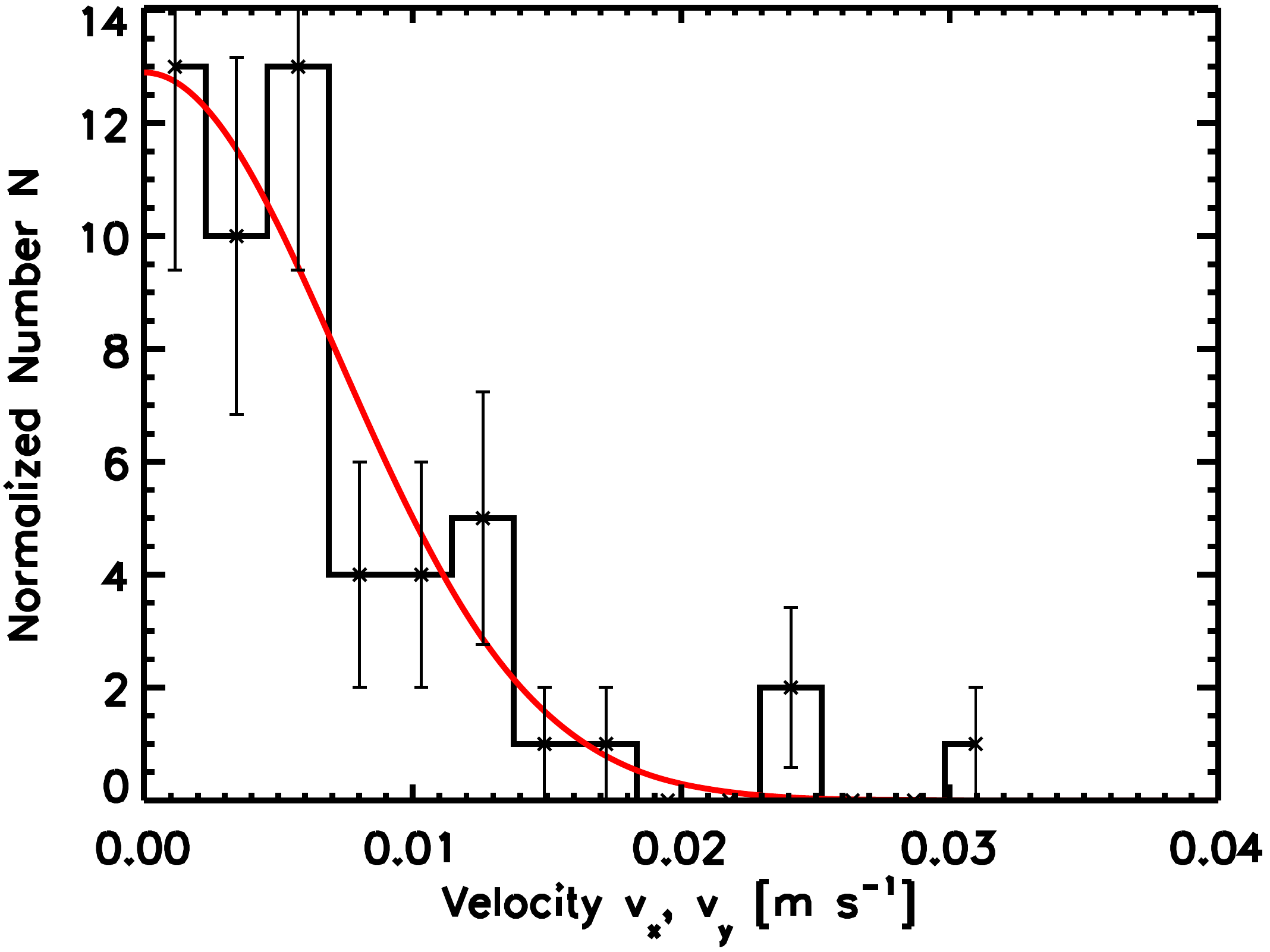}
    \caption{The obtained absolute values of the one-dimensional velocities of the binary collisions (histogram with $1\sigma$ error bars) can be fitted by the positive branch of a Maxwell-Boltzmann distribution (solid curve).}
    \label{f_maxwell2}
\end{figure}

In the drop tower experiments with the particle ensemble we have no information about the three-dimensional motion of the particles, so that we are forced to restrict ourselves on the measured two-dimensional projection of the velocity vector before and after the collisions. Splitting the velocity vectors into two one-dimensional components, we can describe the distribution of the absolute velocity values by a one-dimensional Maxwell-Boltzmann distribution function (see right curve in Fig. \ref{f_maxwell}, Fig. \ref{f_maxwell2}) with a fixed coefficient of restitution of $\overline{\varepsilon}= 0.77$. Using a range of coefficients of restitution, i.e. $\varepsilon=0.2\ldots 0.9$, results in a very similar velocity distribution function (left curve in Fig. \ref{f_maxwell}). The shaded areas in Fig. \ref{f_maxwell} are the result of a Monte Carlo simulation of a one-dimensional Maxwell-Boltzmann velocity distribution and denote the $2\sigma$-range.

\section{Conclusion}\label{s_conclusion}

In this work, we present first results of two novel experimental methods to study the collision properties of individual macroscopic ice bodies, allowing free encounters without limitations present in previous experiments.\par

The parabolic-flight setup is suitable for pair-collisions at relative velocities $\ge 6\,\mathrm{cm\,s^{-1}}$ and near-central impact angles. For normalized impact parameters of $b/R=0 - 0.5$ we could not observe a correlation between the coefficient of restitution and the impact parameter nor with the relative impact velocity of ice particles. The fit of a uniform distribution of the coefficient of restitution resulted in a mean value of $\overline{\varepsilon} = 0.45$, and a range from $\varepsilon = 0.06$ to $0.84$. The experiments with equal-sized and equal-shaped glass particles with collision velocities of a few $\mathrm{cm\,s^{-1}}$ and normal impacts also showed a wide distribution of coefficients of restitution between $\varepsilon = 0.35$ and $\varepsilon = 0.95$. While for the ice particles one could argue that surface frost might be responsible for the scatter in the coefficient of restitution this clearly plays no role for the glass beads. Thus, we expect not a single value for the coefficient of restitution for the collisions in Saturn's rings, but a more or less wide distribution around a mean value. Thus, also very elastic and very inelastic collisions concurrently take place in Saturn's rings. The experiments performed by \citet{bridges_et_al1984Nature} and \citet{hatzes_et_al1988MNRAS} -- although investigating central collisions at lower velocities than we did -- showed only a weak velocity dependence of $\varepsilon$ above $1\,\mathrm{cm\,s^{-1}}$ with values of $\varepsilon \approx 0.2$ and $\varepsilon \approx 0.5$, respectively, and thereby enclose the results obtained from our parabolic flight experiments. The experiments by \citet{dilley_crawford1996JGR} show a decrease of the coefficient of restitution with decreasing projectile mass and with increasing impact velocities. These results do not match those obtained from our parabolic flight experiments. Recent works by \citet{grasselli_et_al2009EL} and \citet{sorace_et_al2009MRC} presenting the investigation of granular cooling and low-velocity binary collisions of mm-sized spherical particles, respectively, report a decrease of the coefficients of restitution with decreasing impact velocities. This finding is supported by the collision model of \citet{brilliantov_et_al2007PRE}. This means that it is unlikely that the deviation of our data points from Haff's law (see Fig. \ref{f_haff}a) for $t>7\,\mathrm{s}$ is due to an increase in the coefficient of restitution. Clustering seems to be a plausible explanation, but needs further exploration.\par

Recent N-body simulations by \citet{porco_et_al2008AJ} were carried out to create a model of Saturn's A and C ring. This model was used as input for numerical light-scattering studies trying to match the photometric properties observed by Cassini. The authors found that to reproduce their observations with their
numerical codes, they required a velocity-dependent coefficient of
restitution of $\sim 3.5$ times lower than that found by \citet{bridges_et_al1984Nature} and 5 times lower than those obtained by \citet{supulver_et_al1995Icarus}. These results could not be reproduced by our experiments. However, the presence of surface frost or regolith layers will alter the collisional properties towards more inelastic behavior. A similar trend is known from impact experiments of cm-sized spheres into lunar and martian regolith analog material \citep{colwell_taylor1999Icarus, colwell2003Icarus} and low-velocity collision experiments of high-porosity dust aggregates \citep[see e.g. ][]{blum_muench1993Icarus, heisselmann_et_al2007IAC, langkowski_et_al2008ApJ}.\par

In a future microgravity campaign the investigation of ice collisions will be extended to impacts onto a solid ice target and at arbitrary impact angles using a slightly modified parabolic flight setup.\par

Using the setup developed as a feasibility study for drop tower experiments, we were able to achieve collisions between cm-sized glass beads at velocities well below $1\,\mathrm{cm\,s^{-1}}$, which could be fitted assuming a mean coefficient of restitution of $\varepsilon = 0.64$. The obtained relative velocities are realistic parameters for simulating the collision processes in Saturn's dense main rings \citep{esposito2002RPPh} so that future drop tower experiments using a cryogenic setup and ensembles of water ice samples will provide further insights into the collision dynamics of icy bodies. These experiments will also incorporate beam splitter optics to provide three-dimensional collision information, which proved to be of utmost importance for the analysis.\par

The observed collisional properties, like the coefficient of restitution $\varepsilon$, are crucial parameters for theoretical and numerical studies \citep{salo_et_al2001Icarus, schmidt_et_al2001Icarus} aiming to explain observed structures, instabilities and overstabilities in Saturn's dense rings. Additionally, the experiments can be utilized to determine the mechanism of energy equipartition within an ensemble of interacting macroscopic grains. For the explanation of the extremely confined ring thickness and the means to counteract perturbations and excitation by dissipation of kinetic energy, more detailed investigations shall be conducted to study the influence of regolith-covered surfaces as well as the accretion and chip-off of debris in collisions of icy bodies. If ring particle collisions are dominated by the particles' surface properties and not by their icy cores, a regolith layer will clearly reduce the coefficient of restitution considerably \citep{heisselmann_et_al2007IAC, blum_muench1993Icarus, langkowski_et_al2008ApJ}.\par

\section{Future Work}
In future experiments using a setup similar to the one used for the described drop tower experiment we will collide ensembles of cm-sized ice particles at cryogenic temperatures. Additionally, low-velocity binary collisions of icy samples will be studied in our laboratory mini drop tower. Both experiments will include stereo optics to gain three-dimensional collision information giving insight into the coefficient of restitution as a function of the impact parameter and the vectorial collision velocity as well as studying the equipartition of translation and rotational energy. The sample properties will also be varied to investigate the effects of surface properties (frost or regolith layers) and shape.

\section*{Acknowledgements}
We are greatful to M. Böttger, S. Buschschlüter, S. Kothe, and M. Thede for their contribution to the development and conduction of the drop tower experiments.\par

We owe Bob Dawson a debt of gratitude for developing, building and improving components of the parabolic flight apparatus and for his invaluable support during the parabolic flight campaign. We also acknowledge D. Salter, G. Chaparro and G. van der Wolk for their dedicated work during the construction of the parabolic flight experiment and R. Weidling for his efforts during the image processing.

This work was co-funded by the European Space Agency (ESA), the Scottish Universities Physics Alliance (SUPA) Astrobiology Equipment Fund, and the German Space Agency (DLR) under grant No. 50 WM 0636.

\bibliography{Literatur}

\clearpage

\noindent \small \textbf{Movie caption:} High-speed image sequence (115~fps) of an ensemble of cm-sized glass spheres colliding in microgravity. The velocity decrease due to inelastic collisions is in agreement with Haff's cooling law for granular `fluids' \citep{haff1983JFM} with a mean coefficient of restitution of $\varepsilon =0.64$. The black dots on the particles are marks for later use in the determination of particle rotation.

\end{document}